\input{aipcheck}

\documentclass[
    ,final            
  ]
  {aipproc}

\layoutstyle{6x9}

\newcommand{\msun}{M$_{\odot}$}

\newcommand{\apj}{ApJ}
\newcommand{\apjl}{ApJ}
\newcommand{\aj}{AJ}
\newcommand{\apjs}{ApJS}

\newcommand{\planss}{Planet. \& Sp. Sci.}

\begin{document}

\title{Initial Conditions of Planet Formation:\\
Lifetimes of Primordial Disks}


\author{Eric E. Mamajek}{
  address={Department of Physics \& Astronomy, University of Rochester, Rochester, NY 14627-0171}
}

\begin{abstract}
The statistical properties of circumstellar disks around young stars
are important for constraining theoretical models for the formation
and early evolution of planetary systems. In this brief review, I
survey the literature related to ground-based and {\it Spitzer}-based
infrared (IR) studies of young stellar clusters, with particular emphasis on
tracing the evolution of primordial (``protoplanetary'') disks through
spectroscopic and photometric diagnostics. The available data
demonstrate that the fraction of young stars with optically thick
primordial disks and/or those which show spectroscopic evidence for
accretion appears to approximately follow an exponential decay with
characteristic time $\sim$2.5 Myr (half-life $\simeq$ 1.7 Myr). Large
IR surveys of $\sim$2--5 Myr-old stellar samples show that there is
real cluster-by-cluster scatter in the observed disk fractions as a
function of age. Recent {\it Spitzer} surveys have found convincing
evidence that disk evolution varies by stellar mass and environment
(binarity, proximity to massive stars, and cluster density). Perhaps
most significantly for understanding the planeticity of stars, the
disk fraction decay timescale appears to vary by stellar mass, ranging
from $\sim$1 Myr for $>$1.3 M$_{\odot}$ stars to $\sim$3 Myr for
$<$0.08 M$_{\odot}$ brown dwarfs. The exponential decay function may
provide a useful empirical formalism for estimating very rough ages
for YSO populations and for modeling the effects of disk-locking on
the angular momentum of young stars.
\end{abstract}

\maketitle


\section{Introduction}

For the 2nd Subaru International Conference on ``Exoplanets and Disks:
Their Formation and Diversity'', the goal of my introductory talk was
to succinctly summarize some recent observational constraints on the
parameters describing circumstellar disks relevant to planet
formation. Rather than give short shrift to many subtopics within too
few pages, I have elected to focus this manuscript on only one aspect
of disk evolution: {\it primordial disk lifetimes}.  For reviews on
recent results on other aspects of circumstellar disk evolution, I
refer the reader to the other reviews in this volume that were
presented at the Kona meeting: especially those on disk geometry and
coronagraphic polarimetry by M. Perrin, observations of debris disks
by A.  Moro-Mart\'{i}n, dynamical theory of planets, planetesimals,
and dust by M. Wyatt, imaging debris disks by P. Kalas, modeling
circumstellar disks by S. Wolf, spectroscopy of disks by M. Goto,
characterizing gas and dust around main sequence stars by C. Chen, and
dust growth in disks by H. Tanaka. For more exhaustive recent reviews
on circumstellar disks around young stars, I refer the reader to the
book by
\citet{Hartmann98}, and the reviews by
\citet{Beckwith00}, \citet{Hillenbrand02}, \citet{Hillenbrand05},
\citet{Meyer07}, \citet{Monin07},
and \citet{Alexander08}.

Investigations of circumstellar disks and extrasolar planets have
experienced explosive (and seemingly parallel) growth over the past
two decades. Our knowledge of extrasolar planets has been driven
mostly by enhanced techniques of optical spectroscopy (e.g. Doppler
surveys), and more recently, photometry (e.g. transits,
microlensing). Our knowledge regarding circumstellar disks orbiting
young stars has been driven mostly by infrared and millimeter
photometry and imaging, and more recently, spectroscopy (with obvious
contributions to understanding accretion from optical spectroscopy).
Protoplanetary disks set the initial conditions of planet formation,
and to understand the diversity of planetary systems, we need to
understand the physics, chemistry, and evolution of primordial disks.

\section{Primordial Disks}

The existence of primordial disks around young stars was originally
inferred through the spectroscopic evidence for accretion among
classical T Tauri stars ($<$2 \msun\, pre-main sequence stars) and
evidence of circumstellar dust structures, later confirmed to be
geometrically distributed as disks \citep[see review by ][ and
references therein]{Hartmann98}. Classical T Tauri stars have typical
isochronal ages of $<$5 Myr, typical accretion rates of
$\sim$10$^{-7}$--10$^{-9}$ \msun\,yr$^{-1}$
\citep{Gullbring98}, disk masses of $\sim$5
$\times$ 10$^{-3}$ \msun\, ($\sim$0.5 dex dispersion), and median
disk/star mass ratios of $\sim$0.5\% \citep{Andrews05}. Among the
classical T Tauri star populations in star-forming regions are
weak-line T Tauri stars -- pre-MS stars lacking spectroscopic of
accretion and/or photometric evidence for circumstellar dust. The disk
masses inferred for weak-lined T Tauri stars from submillimeter
observations are almost always $<$10$^{-3.5}$
\msun\,
\citep{Andrews05}. Integration of the minimum-mass solar nebula model
over the range of orbital radii for planets in our solar system leads
to a disk mass of $\sim$10$^{-2}$ \msun, \citep[i.e. $\sim$2$\times$
the typical disk mass inferred for classical T Tauri
stars;][]{Desch07}, indicating that by the time low-mass stars evolve
to the weak-line T Tauri phase their disk surface density is likely to
be at least an order of magnitude too low to form planets like Jupiter
and Saturn. By ages of $\sim$10 Myr, samples of typical solar-type
stars appear to have less than $<$tens of M$_{Earth}$ of gas within a
few AU, and $<$few M$_{Earth}$ of circumstellar gas at orbital radii
of $\sim$10-40 AU \citep[][]{Pascucci06}. Early in the protostar's
life, the disk (and protostar) are fed by infall from a vast molecular
envelope associated with cloud cores.  The molecular clouds forming
embedded star clusters appear to disperse within $\sim$3 Myr of star
formation
\citep{Hartmann01}, effectively removing the source of dense gas
feeding primordial circumstellar disks.  The mechanisms for removing
reservoirs of circumstellar gas are numerous, including viscous
accretion radially inward toward the star (with some material
subsequently ejected outward via jet), outward viscous decretion
(through conservation of angular momentum), accretion into planets,
photoevaporation by the central star, or even photoevaporation by a
neighboring star \citep{Hartmann98, Alexander08}. Given the numerous
mechanisms for dispersing primordial disks, it is perhaps unsurprising
that we find that typical disk lifetimes are similar to the brief
timescale when the protostar is immersed in a sea of dense molecular
gas.

While the longevity of the Sun's protoplanetary disk is unknown, there
are weak constraints. Detailed modeling of the geophysical, thermal,
and rotational evolution of Saturn's outermost large, regular
satellite Iapetus by \citet{Castillo-Rogez07} requires that Iapetus
accreted most of its mass within $\sim$3.4--5.4 Myr\footnote{Although
originally quoted as 2.5--5 Myr, the timescale was recently updated by the 
same group \citep{Matson09} using revised heat production values for
$^{26}$Al decay.} after the formation of the solar system \citep[defined by
the formation of Ca-Al-rich inclusions 4,567.2\,$\pm$\,0.6 Myr
ago;][]{Amelin02}. This suggests that Saturn itself had accreted the
majority of its mass
\citep[which is at least $\sim$67--80\% hydrogen \&
helium;][]{Guillot99} from the Sun's protoplanetary disk within
$\sim$3.4--5.4 Myr. Variations of the `Nice model' by
\citet{Thommes02, Thommes03} can form the 
gas giants Jupiter and Saturn through core accretion, form Uranus and
Neptune as `failed' gas giants in the $\sim$5--10 AU zone, and scatter
them to near their current orbital radii -- all within 5 Myr -- for
models where the disk surface density is roughly an order of magnitude
higher than the canonical \citet{Hayashi81} minimum-mass solar nebula.
These results suggest that the formation of giant planets is at least
plausible within the $\sim$10$^{6-7}$ yr lifetime of gas-rich
protoplanetary disks, given the observed physical properties of
observed primordial disks.

\section{Primordial Disk Lifetimes}

Photometric and spectroscopic observations appear to be telling us
that the circumstellar disks of young stars undergo radically
divergent evolutionary paths at a very young age. Early
$\sim$3-4\,$\mu$m surveys suggested \citep{Lada00}, and recent Spitzer
surveys of nearby young stellar groups have confirmed
\citep{Lada06, Carpenter06}, that primordial accretion disks
around lower-mass stars tend to last longer than around higher-mass
stars. However within a given mass range, there appears to be quite a
dispersion in disk lifetimes. Shorter disk lifetimes have been
demonstrated for stars which are higher in mass
\citep[e.g.][]{Carpenter06}, in multiple systems
\citep[e.g.][]{Bouwman06, Cieza09}, and which are in the immediate vicinity
($<$0.5 pc) of O-type stars \citep[e.g.][]{Balog07, Mercer09}, or are
in denser stellar clusters \citep[e.g.][]{Luhman08}. Hence there
appear to be multiple variables affecting primordial disk lifetimes.
Although the well-cited disk survey of \citet{Haisch01} pointed
towards a maximum disk lifetime of 6 Myr, we now know of many
convincing examples of older accretors: multiple examples in the
$\sim$6--10 Myr-old $\eta$ Cha, TW Hya, and 25 Ori groups, the accretor
PDS 66 in Lower Cen Crux (stellar age $\simeq$7--17 Myr), and the
unusual binary StH$\alpha$ 34 ($\sim$8--25 Myr)
\citep[][]{Hillenbrand05}.

\begin{figure}
  \includegraphics[width= 4in]{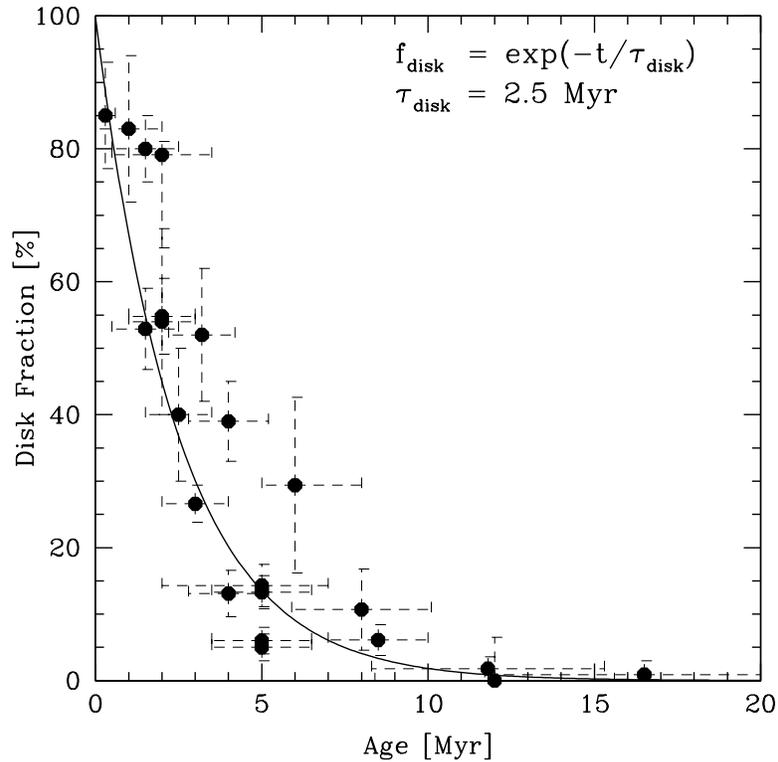} \caption{Age of stellar
  sample vs. fraction of stars with primordial disks (the
  ``Haisch-Lada$^2$'' plot) either through H$\alpha$ emission or
  infrared excess diagnostics. The best fit exponential decay curve is
  plotted with timescale $\tau_{disk}$ = 2.5 Myr. Disk fraction data
  are plotted for (in age order) NGC 2024 \citep[0.3
  Myr;][]{Haisch00}, NGC 1333
\citep[1 Myr;][]{Gutermuth08}, Taurus
\citep[1.5 Myr;][]{Barrado03}, Orion Nebula Cluster \citep[1.5 Myr;][]{Haisch01},
NGC 7129 \citep[2 Myr;][]{Gutermuth04}, NGC 2068/71 \citep[2
Myr;][]{Flaherty08}, Cha I \citep[2.6 Myr;][]{Luhman04, Luhman08}, IC
348
\citep[2.5 Myr;][]{Lada06}, $\sigma$ Ori \citep[3 Myr;][]{Hernandez07},
NGC 2264 \citep[3.2 Myr;][]{Haisch01}, Tr 37 \citep[4.2
Myr;][]{Sicilia-Aguilar06}, Ori OB1b \citep[4 Myr;][]{Hernandez07},
Upper Sco \citep[5 Myr;][]{Carpenter06}, NGC 2362 \citep[5
Myr;][]{Dahm07}, $\gamma$ Vel \citep[5 Myr;][]{Hernandez08}, $\lambda$ Ori
\citep[5 Myr;][]{Barrado07}, $\eta$
Cha \citep[6 Myr;][]{Megeath05}, TW Hya \citep[8 Myr;][]{Barrado03},
25 Ori \citep[8 Myr;][]{Hernandez07, Hernandez08}, NGC 7160
\citep[11.8 Myr;][]{Sicilia-Aguilar06}, $\beta$ Pic 
\citep[12 Myr;][]{Jayawardhana06}, UCL/LCC \citep[16 Myr;][]{Mamajek02}.
\label{fig:diskfrac}}
\end{figure}

A modern version of the ``Haisch-Lada$^2$ plot'' (disk fraction
vs. sample age) is shown in Fig. \ref{fig:diskfrac}. While the
observational definitions of what constitutes a star with a probable
primordial accretion disk can vary slightly from study to study, this
plot represents a best effort to summarize the situation with the data
available. As demonstrated by \citet{Haisch00} from the sample of
known T Tauri stars in Taurus \citep{Kenyon95}, 3-4\,$\mu$m excesses
trace spectroscopically identified classical T Tauri stars $\sim$100\%
of time, while $\sim$2\,$\mu$m excesses trace classical T Tauri stars
only $\sim$70\% of the time. Similarly, \citet{Silverstone06} detected
3.6\,$\mu$m excess emission solely around previously known classical T
Tauri stars among a large sample of FGK-type stars with ages $\sim$3
Myr to $\sim$3 Gyr. With these findings in mind, I consider stars to
be accretors whether they are (1) spectroscopically identified as
classical T Tauri stars (through a H$\alpha$ emission criterion),
and/or (2) through L-band or Spitzer 3.6\,$\mu$m excess, and/or (3)
their emission beyond 3.6\,$\mu$m was classified by the authors as
being likely due to primordial disk due to the SED shape or strength
of the IR excess. The nature of some disks is unclear.
\citet{Lada06} and others have identified stars with weak IR excesses
whose nature as stars with either accretion disks of lower optical
depth or simply warm dusty debris disks is at present ambiguous. Given
the rarity of ``transition disks\footnote{A glossary for common
terminology for young stellar objects with disks (the ``diskionary'')
was recently compiled by \citet{Evans09}.}'' (disks with large inner
holes), their inclusion or exclusion is usually within the disk
fraction uncertainties, and will have negligible impact on this
analysis. The fraction of stars in the transition phase has been noted
to be very high in a young sample
\citep[e.g. $\sim$1 Myr CrA;][]{Sicilia-Aguilar08}.
I have not yet attempted to disentangle the effects of stellar mass in
Fig. \ref{fig:diskfrac}, so the reader should simply interpret the
disk fractions as being most representative of the low-mass population
of stars ($<$2
\msun) as they dominate the stellar samples. I have omitted results
for more distant clusters whose disk fraction statistics were
completely dominated by high mass stars ($>$1--2 \msun).

Ages and age uncertainties are taken from the published studies,
however minimum age uncertainties of $\pm$1 Myr or $\pm$30\%
(whichever is greater) were adopted if uncertainties were not
quoted. The usual caveats exist for the ages plotted in
Fig. \ref{fig:diskfrac}. There are significant differences in the ages
estimated using different evolutionary tracks, and even as a function
of mass for a single set of tracks
\citep{Hillenbrand08}, and none of the tracks have consistently 
matched predictions of masses with dynamically-constrained masses over
the stellar mass spectrum.

Fig. \ref{fig:diskfrac} demonstrates that any statements regarding the
lifetimes of primordial disks need to be statistical in
nature. Statements to the effect that ``all disks disappear'' by
$\sim$3 or $\sim$6 Myr are oversimplified assessments. I have
decided to be provocative and plot an exponential function to fit the
data. An exponential is convenient as it has a value of one at age
zero (assuming that all stars are born with disks), trends towards
zero as age increases (all primordial accretion disks eventually
disappear), and one only needs to fit one parameter (the e-folding
timescale $\tau_{disk}$).  When simultaneously minimizing the
residuals in the disk fraction and age (Fig. \ref{fig:diskfrac}), I
arrive at a best fit timescale of $\tau_{disk}$ = 2.5
Myr. Unfortunately the best fit has reduced $\chi^2_{\nu}$ $\simeq$
2.5, suggesting that either (1) the uncertainties in the assessed disk
fractions and ages are significantly underestimated (by factor of
$\sqrt{2.5}$ $\simeq$ 1.6), and/or (2) there are significant
cluster-to-cluster differences in disk fraction decay time, and/or (3)
an exponential function is simply inadequate for fitting the trend. As
cluster-to-cluster variations have been demonstrated -- especially at
age $\sim$5 Myr
\citep{Carpenter06, Balog07, Luhman08, Dahm09}, and cluster ages are
especially uncertain, I suspect that these two factors are the primary
causes of the high $\chi^2$ of the best fit. Removing individual
clusters from the fit varies $\tau_{disk}$ by $<$10\%, which is
probably a reasonable estimate of the precision in $\tau_{disk}$ given
our current knowledge of the ages of these star-forming regions.

Recent Spitzer surveys have quoted disk fractions as a function of
stellar mass in cluster samples. We can now look at these data in a
different way, and combine the various results and quote a single
metric ($\tau_{disk}$) to more concisely summarize the observed trend
and allow comparison between non-coeval samples. The exponential decay
formalism is convenient for calculating e-folding times for various
subsamples where we have sparse data available. If disk fraction
evolves as an exponential decay, and a cluster of a given age is
observed to have a disk fraction that is assumed to lie on that decay
curve, one can estimate the primordial disk decay timescale from that
subsample:

\begin{equation}
\tau_{disk} = -\tau / {\rm ln}(f_{disk})
\label{eqn:taudisk}
\end{equation}

Where $\tau$\, is the age of the sample, and $f_{disk}$ is the
observed disk fraction.  Using this technique, I will estimate
preliminary values of $\tau_{disk}$ for stellar subsamples segregated
by stellar mass.

{\it What is the characteristic timescale for the primordial disks
around brown dwarfs?} Results from early surveys identifying 3-4\,$\mu$m
excesses among small samples of substellar objects hinted that half of
all disks were likely dispersed within $\sim$1--3 Myr 
\citep{Jayawardhana03, Mohanty05}. More recent disk fractions for larger
samples of substellar objects have been quoted in a series of papers
by Luhman and collaborators, notably for IC 348
\citep[$\tau$ = 2.5 Myr; $f_{disk}$ = 42\%;][]{Luhman05_BD}, Cha I
\citep[$\tau$ $\simeq$ 2.5 Myr; $f_{disk}$ = 50\%;][]{Luhman05_BD}, and $\sigma$
Ori \citep[$\tau$ = 3 Myr; $f_{disk}$ $\simeq$ 60\%;][]{Luhman08}. To
this set, I include the disk fraction of substellar objects in Upper
Sco ($\tau$ = 5 Myr) from \citet{Mohanty05}, which ranged from 
$\sim$5--20\% depending on whether the accretors were defined via
spectroscopic or photometric techniques (we assume 12.5\,$\pm$\,7.5 \%).
Note that the individual $f_{disk}$ values have uncertainties of
$\sim$7--20\%, which translate into significant errors in
$\tau_{disk}$, especially for small $f_{disk}$. Using equation
\ref{eqn:taudisk}, these disk fractions and ages translate into decay
timescales of $\tau_{disk}$ $\simeq$ 2.9, 2.9, 5.9, and 2.4 Myr, for
IC 348, Cha I, $\sigma$ Ori, and Upper Sco, respectively. These
results suggest that the typical primordial disk decay timescale for
brown dwarfs is approximately $\tau_{disk}$ $\simeq$ 3 Myr --
i.e. marginally longer than for stellar samples ($\sim$2.5 Myr).

{\it What is the characteristic decay timescale for primordial disks
around high mass stars?} The small disk fractions, and low numbers of
high mass stars in stellar groups, make this surprisingly difficult to
quantify. \citet{Hernandez05} conducted a systematic survey of the
nearest OB associations to quantify the fraction of Herbig Ae/Be stars
($>$2\,\msun). Their results found f$_{disk}$ $\leq$ 5\% for all of
their samples (3-15 Myr), and for their two youngest samples:
$f_{disk}$ = 5.1\,$\pm$\,2.0\% in Ori OB1bc (3.5\,$\pm$\,3 Myr) and
$f_{disk}$ = 4.3\,$\pm$\,1.8\% in Tr 37. Using {\it Spitzer} to survey
the $\sim$2.5 Myr-old IC 348 group, \citet{Lada06} found a disk
fraction of 11\,$\pm$\,6\% among $>$1.3 \msun\, stars.  The
\citet{Hernandez07} survey of the $\sim$3 Myr-old $\sigma$ Ori
clusters does not provide spectral types or masses, but interpolation
of their Fig. 11 and Table 3 suggests a disk fraction of $>$1.3
\msun\, stars of $\sim$10\%, consistent with Lada et al.'s
findings for IC 348. \citep{Carpenter06} found {\it no} evidence for
primordial disks around a sample of 92 BAFG-type ($>$1.3 \msun)
members of $\sim$5 Myr-old Upper Sco (consistent with f$_{disk}$ $<$
4\%; 95\% confidence), however 7/21 ($\sim$35\%; K0--K6) of $\sim$1--1.3
\msun\, stars have primordial disks! The results from these
three {\it Spitzer} surveys are consistent with $\tau_{disk}$ $\simeq$
1.2 Myr for $>$1.3 \msun\, stars. These results are also broadly
consistent with the frequency of $>$2\,\msun\, Herbig Ae/Be stars in
nearby associations \citep{Hernandez05}.

It is possible that $\tau_{disk}$ could be used as a {\it very} coarse
age estimator for multi-wavelength investigations of distant
star-forming regions imaged both in the infrared and in X-rays where
there is an estimate of the young stellar population both with
primordial disks (class O/I/II objects) and without (class III
objects). The mean age of the population would then be approximately
$\tau$ $\simeq$ -ln $f_{disk}$ $\times$ $\tau_{disk}$ where $f_{disk}$
= N$_{disk}$ / (N$_{disk}$ + N$_{no~disk}$) and $\tau_{disk}$ $\sim$
2.5 Myr. This estimate would only provide the coarsest of ages (as we
now have evidence that disk lifetimes are dependent on stellar mass
and environment), however in the quest for useful stellar age
estimators, even the bluntest of diagnostics can be helpful.

In summary, it is clear from {\it Spitzer} surveys that the lifetime
of primordial disks is not only a function of age, but stellar mass,
multiplicity, and proximity to O-type stars. Disk fraction appears to
vary roughly as an exponential decay, with typical timescale
$\tau_{disk}$ $\simeq$ 2.5 Myr. This constant appears to vary from
$\sim$1.2 Myr for $>$1.3 \msun\, stars to $\sim$3 Myr for brown
dwarfs.  Although numerous mechanisms have been posited for depleting
circumstellar disks, we need more observations to better constrain the
disk evolution as a function of these stellar parameters (and for
other untested parameters, e.g. metallicity), and more theoretical
work to model these depletion mechanisms. It is clear that there are
ample future opportunities for observations with the Subaru telescope
to improve our understanding of the formation and early evolution of
planetary systems.


\begin{theacknowledgments}
I thank the SOC for the Subaru conference for inviting me to give this
review talk, and I thank Michael Meyer and Dan Watson for contributing
slides. Thanks also go to Alex Shvonski and Mark Pecaut for commenting
on an early draft.
\end{theacknowledgments}



\bibliographystyle{aipproc}   


\IfFileExists{\jobname.bbl}{}
  {\typeout{}
   \typeout{******************************************}
   \typeout{** Please run "bibtex \jobname" to optain}
   \typeout{** the bibliography and then re-run LaTeX}
   \typeout{** twice to fix the references!}
   \typeout{******************************************}
   \typeout{}
  }

\end{document}